\documentclass[epsfig,graphics,twocolumn,prl,showpacs]{revtex4}%
\usepackage{amsmath}
\usepackage{graphicx}
\usepackage{epsfig,color}
\usepackage{amsfonts}
\usepackage{amssymb}%
\providecommand{\U}[1]{\protect\rule{.1in}{.1in}}

\begin{document}
\title{Nernst effect and disorder in the normal state of high-$T_{c}$ cuprates}
\author{F.\ Rullier-Albenque$^{1}$, R.\ Tourbot$^{1}$, H.\ Alloul$^{2}$,
P.\ Lejay$^{3}$, D.\ Colson$^{1}$, A.\ Forget$^{1}$}
\affiliation{$^{1\text{ }}$SPEC, Orme des Merisiers, CEA, 91191 Gif sur Yvette cedex,
France }
\affiliation{$^{2}$ Laboratoire de Physique des Solides, UMR 8502, Universit\'{e}
Paris-Sud, 91405 Orsay, France }
\affiliation{$^{3}$ CRTBT, CNRS, BP166X, 38042 Grenoble cedex, France}
\date{\today     }

\begin{abstract}
We have studied the influence of disorder induced by electron irradiation on
the Nernst effect in optimally and underdoped YBa$_{2}$Cu$_{3}$O$_{7-\delta}$
single crystals.\ The fluctuation regime above $T_{c}$ expands significantly
with disorder, indicating that the $T_{c}$ decrease is partly due to the
induced loss of phase coherence. In pure crystals the temperature extension of
the Nernst signal is found to be narrow whatever the hole doping, contrary to
data reported in the low-$T_{c}$ cuprate families. Our results show that the
presence of \textquotedblright intrinsic\textquotedblright\ disorder can
explain the enhanced range of Nernst signal found in the pseudogap phase of
the latter compounds.

\end{abstract}

\maketitle

The nature of the pseudogap phase remains a key issue to understand
superconductivity in high-$T_{c}$ cuprates. Among the variety of scenarios
which have been proposed \cite{Timusk}, an important one is to consider the
pseudogap as a precursor to superconductivity, its opening being attributed to
a phase-incoherent pairing, the long range coherence occuring only at $T_{c}$
\cite{Emery1}. An experimental support for this description of the pseudogap
regime has been set out recently by the occurence of a substantial Nernst
signal in the normal state of some underdoped cuprates well above the
transistion temperature $T_{c}$ \cite{Xu,Wang,Wang2}. As this signal is known
to be associated with vortex motion in the mixed state of superconductors, it
has been suggested that these Nernst effet experiments reveal the existence of
vortex-like excitations surviving in the normal state. In any case this Nenst
signal can hardly be explained without invoking superconducting fluctuations
\cite{Wang,Kontani,Ussishkin,Levin,Lee}.\ This has been recently reinforced by
new measurements of a diamagnetic response which has been found to track the
Nernst signal \cite{Wang3}

It has been considered somewhat independently by Emery and Kivelson
\cite{Emery2} that in a sufficiently bad metal classical and quantum phase
fluctuations of the superconducting order parameter can depress $T_{c}$ well
below its mean-field value.\ We have previously shown that the defect induced
decrease of $T_{c}$ could be partly explained within this scenario \cite{RA1}.
One might wonder whether this results as well in a large range of incoherent
phase fluctuations above $T_{c}$.\ In order to test this possibility, we have
undertaken a systematic study on the influence of defects on the Nernst
effect. We have chosen to perform experiments\ in optimally doped YBCO$_{7}%
$\ and underdoped YBCO$_{6.6}$ compounds which are known to be very
homogeneous systems with little intrinsic disorder \ \cite{Bobroff1}. The
controlled introduction of defects has been achieved by using electron
irradiation at low temperature which results in the creation of point defects
such as Cu and O vacancies in the CuO$_{2}$ planes \cite{RA}. We demonstrate
here for the first time that the presence of defects induces the apparition of
a Nernst signal in a large temperature range above $T_{c}$ in both compounds.
This is a strong confirmation that phase fluctuations do play a role in the
decrease of $T_{c}$ induced by disorder. Moreover we find that the onset
temperature of the Nernst effect is not much dependent on the defect content
and remains close to that of the pure system. We shall discuss the
implications of these results on the analysis of the existing data on systems
with lower intrinsic $T_{c}\ $such as LaSrCuO or La-doped Bi2201 \cite{Wang}.

The single crystals used in this study were grown using the standard flux
method. Very small contacts with low resistance ($<0.1\Omega$) were achieved
by evaporating gold pads on the crystals on which gold wires were attached
later with silver epoxy. Subsequent annealings have been performed in order to
obtain crystals with oxygen content \symbol{126}7 and \symbol{126}6.6. The
$T_{c}$ values are defined here as the zero resistance temperatures. The
irradiation were carried out with 2.5MeV electrons in the low temperature
facility of the Van der Graaff accelerator at the LSI (Ecole Polytechnique,
Palaiseau).\ During irradiation, the samples were immersed in liquid H$_{2}$
and the electron flux was limited to 10$^{14}e/cm^{2}/s$ \ to avoid heating of
the samples during irradiation. The thicknesses of the samples (20 to 40 $\mu
m$) are very small compared to the penetration depth of the electrons, which
warrants an homogeneous damage throughout the samples. We report here data
taken on three YBCO$_{7}$ samples\ : a pure one with $T_{c}=92.6$K and two
irradiated ones at different electron fluences with respective $T_{c}=79.5$K
and $48.6$K,\ and four YBCO$_{6.6}$ samples : pure with $T_{c}=57K$ and
irradiated with $T_{c}=45.1$K, $24.2$K and $3$K.

The Nernst signal $E_{y}$ is the the transverse electrical response to a
thermal gradient $\nabla_{x}T//x$ in a presence of a perpendicular magnetic
field $B//z$.\ For the measurements the sample was attached on one end to a
copper block with the other end free.\ The temperature gradient was created
with a small RuO$_{2}$ resistance attached to the free end.\ The measurements
were performed under vacuum ($10^{-2}$ to $10^{-1}$ mbar) and a heater power
ranging from $0.01$ to $0.2$ mW was used to create temperature gradients from
$0.5$ to $0.8$K/mm depending on the temperature of measurement. The thermal
gradient was measured with a differential chromel-constantan thermocouple. The
data were taken at fixed $T$ with magnetic field sweeps from 0 to 8T. At some
given value of the magnetic field, the thermal gradient is removed, which
allows us to subtract offset voltages due to contact misalignement or an
eventual contribution of the wires.

As described by Wang et al. \cite{Wang}, the Nernst coefficient $\nu
=E_{y}/(-\nabla_{x}T)B$ is the contribution of two terms :%
\begin{equation}
\nu=\frac{E_{y}}{(-\nabla_{x}T)B}=\left[  \frac{\alpha_{xy}}{\sigma}%
-S\tan\theta\right]  \frac{1}{B} \label{Eq.1}%
\end{equation}
where $\alpha_{xy}$ is the off-diagonal Peltier conductivity \ ($J_{y}%
=\alpha_{xy}(-\nabla_{x}T)$), $\theta=\sigma_{xy}/\sigma$ is the Hall angle
and $S$ the thermopower. The quantity of interest is the off-diagonal term
$\alpha_{xy}$ which involves the normal-state term $\alpha_{xy}^{n}$ and the
vortex contribution $\alpha_{xy}^{s}$. In order to probe the influence of
disorder on the latter, it is very important to determine as well the
influence of disorder on $S\tan\theta$. We have therefore measured $E_{y}$,
$S$ and $\tan\theta$ separately in each sample by using the same electrodes
for measuring resistivity and Hall effect in one setup and the thermopower and
Nernst coefficients\ in another one.

Let us present first the results obtained on underdoped crystals. Figure 1
shows several curves of the Nernst signal $e_{y}=E_{y}/(-\nabla_{x}T)$ as a
function of magnetic field for the pure crystal.%

\begin{figure}
[ptbh]
\begin{center}
\includegraphics[
natheight=5.291800in,
natwidth=6.153100in,
height=6.8447cm,
width=8.6525cm
]%
{../plot1.wmf}%
\caption{(color online) Nernst signals $e_{y}=E_{y}/\left|  \nabla T\right|  $
versus magnetic field in the pure underdoped YBCO$_{6.6}$ crystal for $T$
ranging from $35$ to $200$K}%
\label{Fig.1}%
\end{center}
\end{figure}

For $T<55$K, $e_{y}$ is zero as long as the magnetic field does not exceed the
''melting'' field $B_{m}(T)$ necessary to depin vortices (around $2$T and $1$T
respectively at $35$ and $45$K).\ Then the rapid increase of $e_{y}$ above
$B_{m}$ reflects the motion of vortices induced by the thermal gradient. As
$T$ is increased across $T_{c}$ the Nernst signal initially drops rapidly
(curves at $55$ and $58$K) and then decreases gradually approaching a straight
line with negative slope. This behaviour is clearly displayed in Fig.2a in
which we have plotted the $T$ variation of the Nernst coefficient $\nu$
determined as the initial slope of $e_{y}$ versus $B$. This negative
contribution which has been previously observed near $T_{c}$ \cite{Ong} is
found to display a minimum at $85K$ and to vanish around $200K$. This
behaviour which is quite different from that observed in the other underdoped
cuprates, will be seen below to result naturally from the high value of
$S\tan\theta/B$ in this clean system.%

\begin{figure}
[ptb]
\begin{center}
\includegraphics[
natheight=8.777800in,
natwidth=6.555300in,
height=9.887cm,
width=7.8485cm
]%
{../plot2.wmf}%
\caption{(color online) Temperature evolution for the pure and two irradiated
underdoped YBCO$_{6.6}$ samples of : (a) the Nernst coefficient determined by
the initial slope of $e_{y}$ versus $B$ ( the $T$ dependence of \ $S\tan
\theta$ is also plotted for the pure and the most irradiated samples), (b)
$\alpha_{xy}/\sigma B$ determined from Eq.1 and (c) the resistivity. The
arrows in (b) indicate the onset temperature of the vortex Nernst
contribution.}%
\label{Fig.2}%
\end{center}
\end{figure}

The effect of electron irradiation is recalled in Fig. 2c where the
resistivity curves $\rho(T)$ are plotted for the pure and two irradiated
samples.\ As reported previously \cite{RA1}, Matthiessen's rule is well obeyed
at high $T$, which indicates that the hole doping of the CuO$_{2}$ planes and
the pseudogap temperature T$^{\ast}$ are not significantly modified
\cite{Alloul}. The initial parts of the low $T$ upturn of $\rho(T)$ have been
associated with a Kondo-like spin flip scattering \cite{RA2}. As for the
Nernst coefficient one observes in Fig.2a that the pronounced minimum which is
present for the pure sample is smoothed out by the introduction of disorder.
We have reported on the same graph the temperature dependence of $S\tan
\theta/B$ for the pure and the most irradiated sample. In the pure sample both
$S$ and $\tan\theta$ being quite large \cite{footnote1}, $-S\tan\theta/B$
dominates in Eq.\ref{Eq.1}. Such a negative value of $\nu$ has been
predicted\ theoretically in the framework of the Boltzmann theory by taking
into account the role of\ the Fermi surface shape at two dimensions
\cite{Oganes}. When increasing the defect content $x$, we find that $S$ varies
slightly while $\tan\theta/B$ decreases roughly as $1/x$ \cite{footnote2},
resulting in a decrease of $S\tan\theta/B$ when $T_{c}$ decreases. The $T$
variation of the total off-diagonal Peltier term\ $\alpha_{xy}/\sigma B$
obtained by combining the data for $S\tan\theta$ and $\nu$ is plotted in
Fig.2-b. In all samples the normal state contribution $\alpha_{xy}^{n}$
presents a broad peak around 110K and then decreases with temperature.\ Such a
behavior has been quite generally observed in underdoped cuprates \cite{Wang}
and might therefore be characteristic of the normal state quasiparticles. As
$\alpha_{xy}^{n}/\sigma B$ corresponds to a carrier-entropy current one
expects that it should decrease to zero at $T\rightarrow0$.\ Therefore it
seems legitimate to interpret any deviation from this tendency as a
manifestation of a vortex contribution. We have thus indicated by the arrows
in Fig.2b the best estimate of the onset temperature $T^{\nu}$ of the
superconducting contribution. This determination leads in fact to values which
nearly coincide with those of the minimum of the Nernst coefficient. Two
important results can be deduced from this plot.\ First it is clearly seen
that the onset temperature does not exceed 85K in pure YBCO$_{6.6}$, showing
that the \textit{fluctuation regime is quite narrow} ($\thicksim25$K) in this
compound despite the fact that the pseudogap temperature $T^{\ast}$ is
$\gtrsim$ $300$K whatever the experimental probe \cite{Tallon}. Second we find
that $T^{\nu}$\textit{\ is nearly the same} \textit{for all the samples} while
$T_{c}$ has been decreased down to $5$K by irradiation.\ This is a strong
indication that the presence of defects plays a prominent role in the
observation of a Nernst signal in the normal state of these samples.

As for YBCO$_{7}$, the Nernst coefficients are reported in Fig.3 for the three
samples studied. In the pure crystal the magnitude of the negative value of
$\nu$ is much smaller than in the underdoped case. This results from the fact
that $S\tan\theta/B$ is also smaller \cite{footnote2} as shown by the
decomposition displayed in the inset of Fig.3. This $S\tan\theta/B$ term
varies very little with defect content and the estimate of $T^{\nu}$ is about
the same whether we use the raw data for $\nu$ or the corrected values
$\alpha_{xy}/\sigma B$. For all the samples the drop of the Nernst signal is
very rapid at $T_{c}$ but while it vanishes at $\symbol{126}10$K above $T_{c}$
in the pure sample, it persists up to \ $\symbol{126}85K$, that is to say
$35$K above $T_{c}$, in the most irradiated one. The fairly narrow fluctuation
range found in the pure sample is similar to the one deduced from the
paraconductivity in the $\rho(T)$ curves.%
\begin{figure}
[ptb]
\begin{center}
\includegraphics[
trim=0.000000in 0.000000in -1.136075in 0.000000in,
natheight=5.097200in,
natwidth=5.069500in,
height=6.5833cm,
width=8.0001cm
]%
{fig3-rev.wmf}%
\caption{(color online) The Nernst coefficient $\nu$ is plotted versus
temperature for pure and irradiated single crystals of YBCO$_{7}$. The values
of T$_{c}$ and T$^{\nu}$ are respectively indicated by dashed and full arrows.
The $T$ dependences of $\nu$ ( circles), $S\tan\theta/B$ ( diamonds) and
$\alpha_{xy}/\sigma B$ (squares) are shown in the inset for the pure crystal.}%
\label{Fig.3}%
\end{center}
\end{figure}

In order to compare results obtained on YBCO$_{7}$ and YBCO$_{6.6}$ we have
reported in Fig.4 the values of $T^{\nu}$ as a function $T_{c}$ for the
different samples. In both compounds we observe that the Nernst signal extends
in a larger temperature range when decreasing $T_{c}$. Let us point out that
this effect corresponds to very small values of the vortex Nernst signal as
one can see that the temperature corresponding to a value of the vortex Nernst
signal of $30nV$/KT nearly follows the $T_{c}$ decrease.

These results clearly show that superconducting fluctuations survive in the
normal state of \textit{both} optimally doped and underdoped YBCO when $T_{c}
$ is decreased by the introduction of disorder. As $T^{\nu}$ can be considered
as the characteristic temperature below which local pairing remains
significant, the $T_{c}$ decrease induced by disorder can only be explained by
taking into account both phase fluctuations and pair-breaking effects. This
gives strong support to our previous interpretation of \emph{\ }the quasi
linear decrease of $T_{c}$ with defect content which is observed down to
$T_{c}=0$ \cite{RA1}. It is worth mentioning here that the role of quantum
phase fluctuations has also been invoked to explain the Nernst effect observed
in the normal state of low $T_{c}$ cuprates when superconductivity is
suppressed by magnetic fields \cite{Capan,Ikeda}.%
\begin{figure}
[ptb]
\begin{center}
\includegraphics[
natheight=5.486400in,
natwidth=5.944700in,
height=8.0309cm,
width=7.027cm
]%
{fig4rev.wmf}%
\caption{(color online) The values of $T^{\nu}$ are plotted versus $T_{c}$
together with the $T$ values corresponding to vortex Nernst contributions of
$10$ and $30n$V/KT for YBCO$_{7}$ (empty symbols) and YBCO$_{6.6}$ (closed
symbols). These data are compared to the temperature ranges of the Nernst
signal measured in ''pure'' single crystals of Bi$_{2}$Sr$_{2-y}$La$_{y}%
$CuO$_{6}$ with $y=0.4$ and $y=0.5$ \cite{Wang}.}%
\label{Fig.4}%
\end{center}
\end{figure}

One striking point which can be seen here is the small range of
superconducting fluctuations observed in the pure YBCO$_{6.6}$ and YBCO$_{7}$
compounds.\ This is much smaller than the corresponding observations done in
other ''pure'' cuprates such as LaSrCuO or La-doped Bi2201 \cite{Xu,Wang}. Let
us recall here that the presence of extended vortex fluctuations in these
underdoped cuprates has been\ invoked as a strong indication that d-wave
superconductivity is closely connected to the pseudogap state
\cite{Wang,Wang3}.\ Our results show that \textit{this argument fails in pure
underdoped YBCO}$_{6.6}$, suggesting that the energy scales of T$^{\nu}$ and
T$^{\ast}$ are not connected. We can even see that in these clean systems
$T^{\nu}$ and $T_{c}$ increase with increasing hole doping while $T^{\ast}$
definitely decreases.

It has been previously suggested that the low $T_{c}$ in some cuprate families
could be due to the presence of intrinsic defects as deduced from the analysis
of $^{17}O$ NMR data \cite{Bobroff2}. One can wonder whether this might also
explain the magnitude of the Nernst effects.\ We have therefore compared in
Fig.4 the temperature extensions of the Nernst signal of our irradiated
samples with those obtained in the Bi$_{2}$Sr$_{2-y}$La$_{y}$CuO$_{6}$ family
\cite{Wang} for $y=0.4$ ($T_{c}\thicksim36K$) which corresponds to optimal
doping and for $y=0.5$ ($T_{c}\thicksim29K$ ) with T$^{\ast}\thicksim$ $300K$
comparable to that of YBCO$_{6.6}$ \cite{Hanaki}. The quite good agreement
between the ranges of vortex Nernst signal found in these different samples
indicates that the ''intrinsic'' disorder in Bi$_{2}$Sr$_{2-y}$La$_{y}%
$CuO$_{6}$ could also be responsible for the enhanced Nernst signal. Indeed
cation disorder on the Sr site has been recently identified
\cite{Eisaki,Fujita}. It is then natural to conclude that the ''anomalously''
high values of $T^{\nu}$with respect to $T_{c}$ found in La-doped Bi2201 are
indicative of the values of $T_{c}$ that these materials should display if
they were grown without local inhomogeneities. Such a conclusion might as well
apply to the various cuprate families \cite{Bi-2212} and is reinforced by the
fact that the maximum of $T^{\nu}$ is in most cases of the order of 100K
\cite{Wang}. Moreover our results reveal that defects induce more phase
fluctuations in the underdoped phase than for optimal doping, which might be
due to lower phase stiffness and less efficient screening. It is therefore our
opinion that the link between the large range of Nernst signal and the
pseudogap phase has to be found in priority in the occurence of defects and in
the large sensitivity to disorder of the superconducting-pseudogap phase.


\begin{thebibliography}{99}
\bibitem {footnote1}The values of $S$ and of $\tan\theta$ measured in this
sample are quite comparable to those reported in the literature for underdoped
YBCO$_{6.6}$ \cite{Cooper}
\bibitem {footnote2}Hall effect measurements on Zn-subsituted or electron
irradiated $YBCO_{7}$ crystals have shown that $\cot\theta$ increases linearly
with the defect content $x$ \cite{Chien, Legris}.
\bibitem {footnote3}The thermoelectric power $S$ is very small ($\leq2\mu V/K
$) in this optimally doped crystal.
\end{thebibliography}
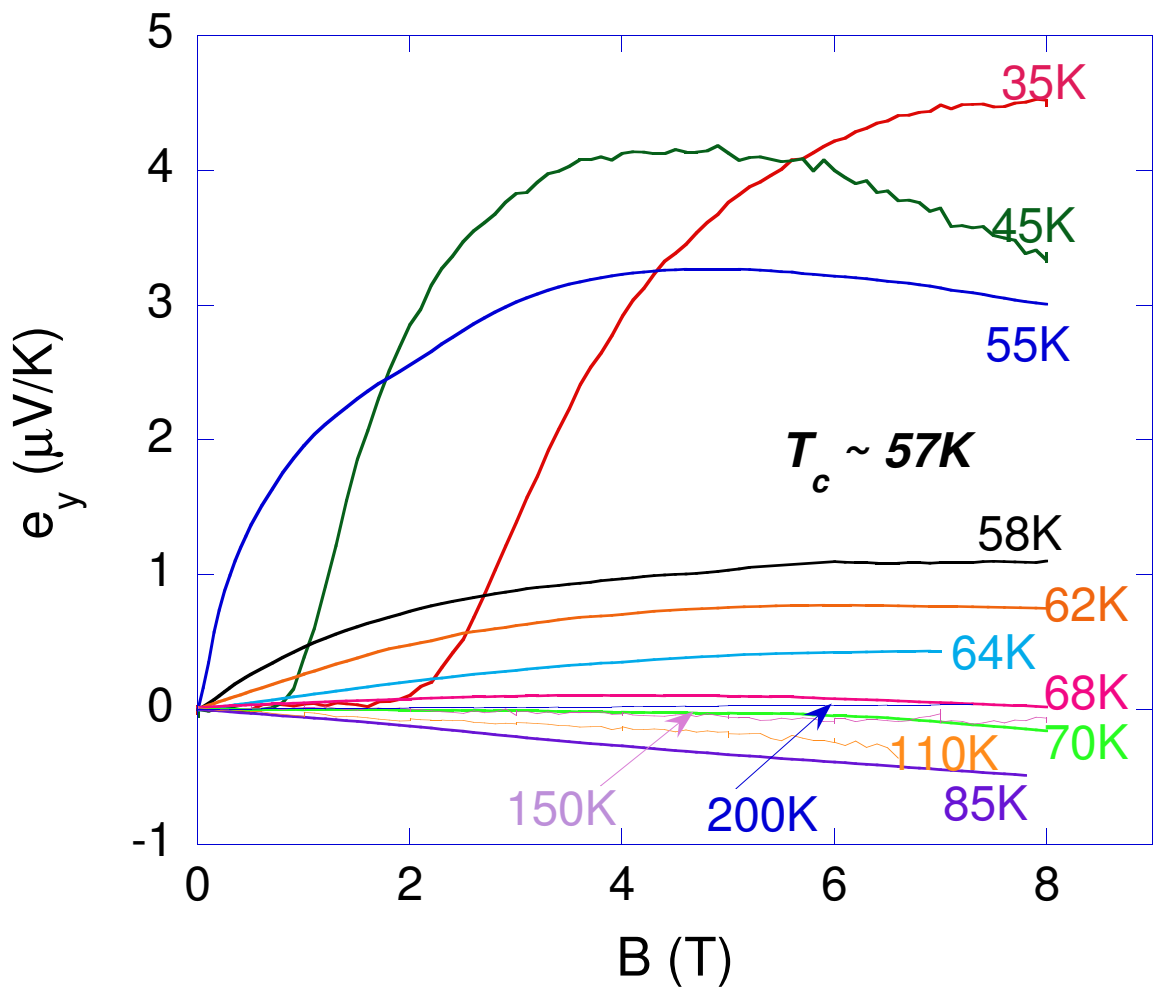
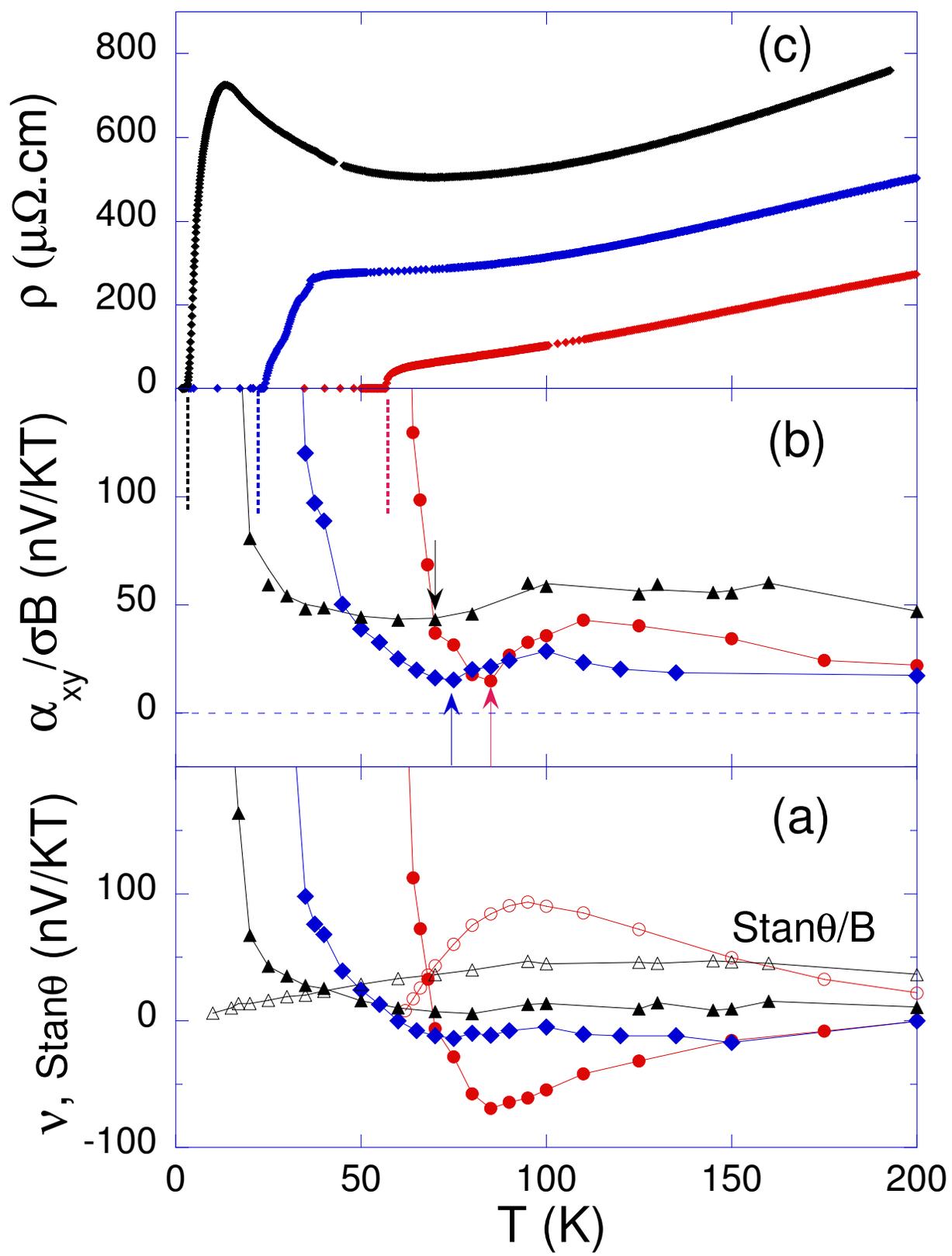
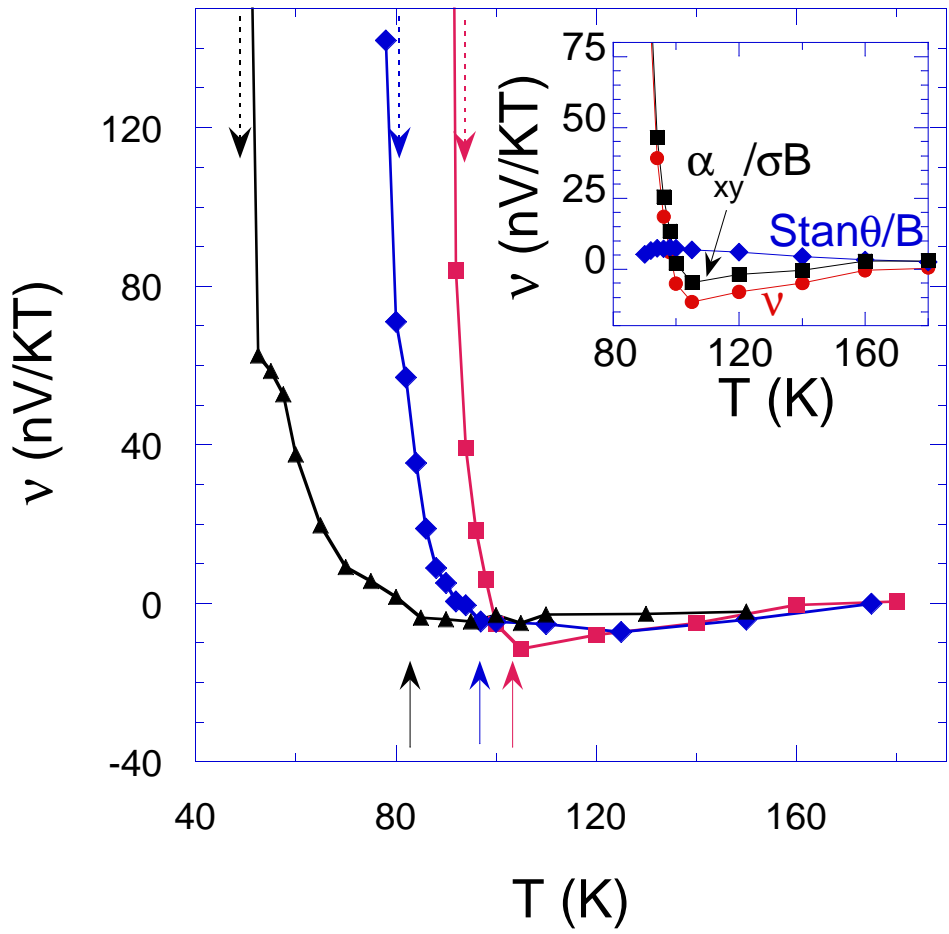
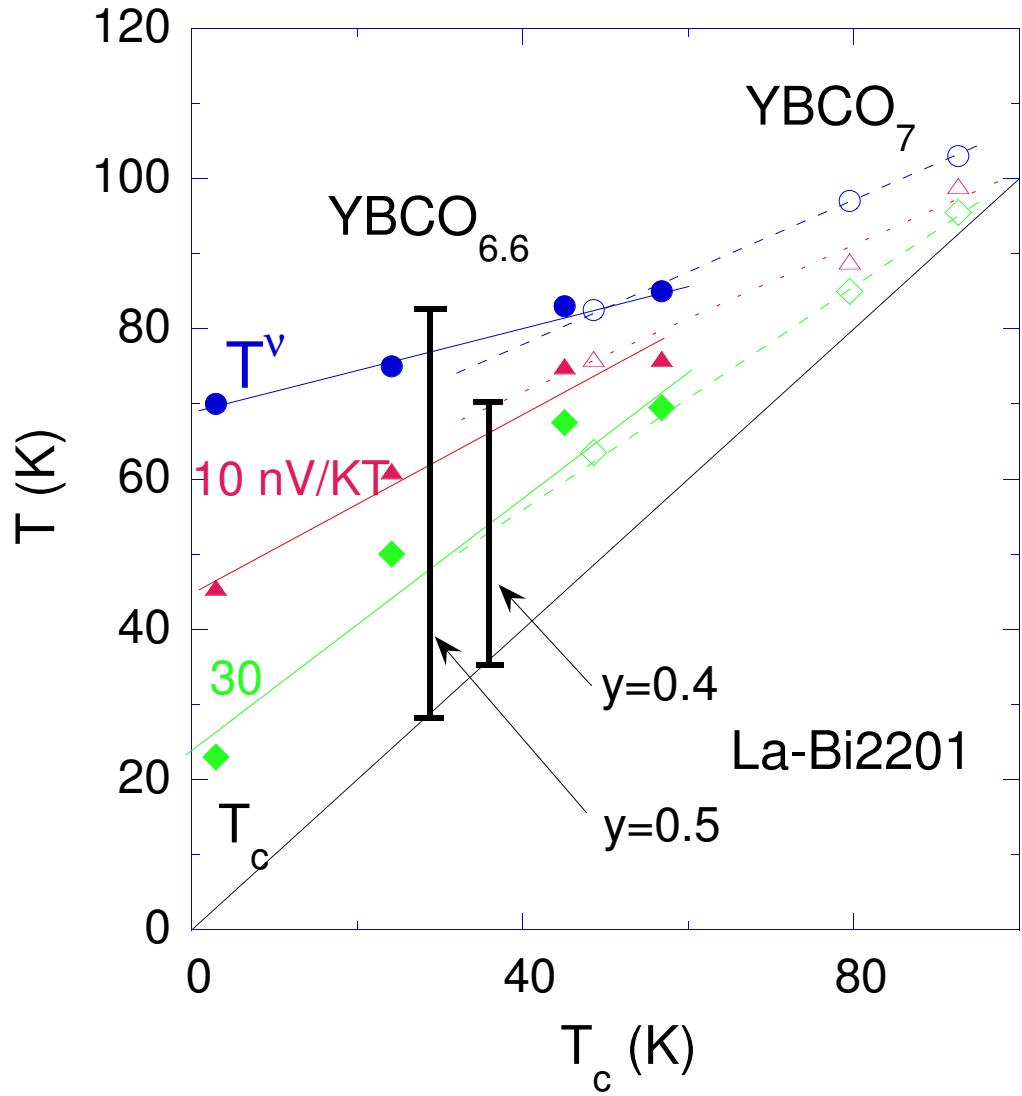
\end{document}